\documentclass[preprint,showpacs,aps]{revtex4}

\usepackage{amsmath}
\usepackage{dcolumn}
\usepackage{verbatim}

\begin{document}
\title{Isospin relations for four nucleons in a single $j$ shell}

\author{L. Zamick and A. Escuderos}

\affiliation{Department of Physics and Astronomy, Rutgers University,
Piscataway, New Jersey USA 08854}

\date{\today}

\begin{abstract}
We had previously used techniques involving isospin to count the number of states
for three identical fermions in a single $j$ shell with total angular momentum
$I=j$. We generalize this to all $I$, but the main thrust of this work is to
consider now a 4-fermion system. As before, one evaluates the eigenvalues of the
Hamiltonian $\sum_{i<j}[a+b t(i) \cdot t(j)]$ both from an isospin point of view
and an angular momentum point of view. In the 4-particle case, we get a more
limited result than in the 3-particle case, namely the number of $T=0$ states
minus twice the number of $T=2$ states, all of a given angular momentum $I$.
\end{abstract}
\maketitle

\section{Introduction}

It is our aim in this work to consider the number of states of a given angular
momentum and isospin for a 4-particle system. We previously considered a system
of three nucleons in a single $j$ shell~\cite{ze05}. By  using isospin
considerations, we were able to obtain a formula for the number  of states with
total angular momentum $j$, although it is easy to generalize this to any
angular momentum $I$ for the 3-particle sytem. We did this by considering the 
expectation value of the simplified interaction $V=\sum_{i<j}[a+b t(i) \cdot
t(j)]$, where  $a$ and $b$ are constant, for a system of two protons and one
neutron in a  single $j$ shell. If we do not choose to use isospin degrees of
freedom (and we do not have to!), the basis states are of the form
\begin{equation}
\left[ \left[ j_\pi (1) j_\pi (2) \right]^{J_P} j_\nu (3) \right]^I ~.
\end{equation}
But we will find it useful to construct explicitly a wave function which is 
antisymmetric in space, spin, and isospin
\begin{equation}
\Psi = \frac{1}{\sqrt{3}} \left( 1 - P_{13} - P_{23} \right) \left[ \left[
j(1) j(2) \right]^{J_P} j(3) \right]^I p(1) p(2) n(3) ~.
\end{equation}

If we focus on the isospin variables, the eigenvalues for $A$ valence nucleons 
are
\begin{equation}
\langle V \rangle = \frac{a}{2} A(A-1) + \frac{b}{2} T(T+1) - \frac{3}{8} Ab ~,
\end{equation}
which, for $A=3$, becomes
\begin{equation}
\langle V \rangle = 3a + \frac{b}{2} T(T+1) - \frac{9}{8} b ~. \label{v-iso3}
\end{equation}

From the angular momentum point of view, it was noted~\cite{ze05} that if we 
set up the Hamiltonian $\langle [J'_P j]^I | V | [J_P 
j]^I \rangle$ and take the trace, we find
\begin{equation}
\text{tr} [J_P] = \sum_{J_P \text{ \footnotesize  even}} \left[ \left( 3a - 
\frac{b}{4} \right) + b (2J_P +1)
\begin{Bmatrix}
j & j & J_P \\
I & j & J_P
\end{Bmatrix}
\right] ~. \label{tracejp}
\end{equation}

If, in eq.~\eqref{v-iso3}, we choose $a$ and $b$ so that $\langle V \rangle = 0$
for $T=1/2$ and $\langle V \rangle = 1$ for $T=3/2$, we find $a=1/6$, $b=2/3$.
With this choice, $\text{tr} [J_P]$ becomes the number of $T=3/2$ states of
angular momentum $I$, which is the same as the number of states of angular
momentum $I$ for a system of three identical particles, namely
\begin{equation}
\frac{1}{3} \sum_{J_P \text{ \footnotesize even}} \left( 1 + 2 (2 J_P +1)
\begin{Bmatrix}
j & j & J_P \\ I & j & J_P
\end{Bmatrix} \right) ~.
\end{equation}
From this we learned that, even though we are dealing with a system of two 
protons and one neutron, we obtain a result relevant to three identical 
nucleons.

In the next section, we will consider four nucleons (two protons and two 
neutrons) and see what information we can learn from taking the expectation 
value of $V$ for this case.

\section{A system of two protons and two neutrons in a single $j$ shell}

We consider a system of two protons coupled to angular momentum $J_P$ and two 
neutrons coupled to $J_N$, and then $J_P$ and $J_N$ coupled to a final angular 
momentum $I$:
\begin{equation}
\left[ \left[ j(1) j(2) \right]^{J_P} \left[ j(3) j(4) \right]^{J_N} \right]^I.
\end{equation}
Here $J_P$ and $J_N$ must be {\it even} in order for the 2-proton (2-neutron) 
wave function to be antisymmetric. We can also include isospin coordinates $p$ 
and $n$ and then construct a wave function that satisfies the generalized Pauli
Principle, that is to say, the wave function is antisymmetric in space, spin, 
and isospin
\begin{equation}
\Psi = \frac{1}{\sqrt{5}} (1 - P_{13} - P_{14} - P_{23} - P_{24} ) 
\left[ \left[ j(1) j(2) \right]^{J_P} \left[ j(3) j(4) \right]^{J_N} \right]^I 
p(1) p(2) n(3) n(4) ~.
\end{equation}

We now want the expectation value of the simplified interaction $V(ij)=a+b t(i)
\cdot t(j)$, where $a$ and $b$ are constants. We can do this both from isospin 
considerations (easy) and from angular momentum algebra (more difficult). Using
the isospin method, we have
\begin{equation}
\left\langle \sum_{i<j} V(i j) \right\rangle = \frac{a}{2} A(A-1) + 
\frac{b}{2} T(T+1) - \frac{3}{8} A b ~,
\end{equation}
where $A$ is the number of valence nucleons that are present. For $A=4$, we get
$6a-3b/2+bT(T+1)/2$.

From the angular momentum point of view, there is more work to do:
\begin{eqnarray}
\left\langle \sum_{i<j} V(ij) \right\rangle & = & \left\langle \Big(1-P_{13} -
P_{14} - P_{23} - P_{24}\Big) \left[ \left[ j(1) j(2) \right]^{J_P} \left[ j(3)
j(4) \right]^{J_N} \right]^I p(1) p(2) n(3) n(4) \Big| \right. \nonumber \\
 & & \left. \sum_{i,j} \Big[a+b t(i) \cdot t(j)\Big] \Big| \left[ \left[ j(1) 
j(2) \right]^{J_P} \left[ j(3) j(4) \right]^{J_N} \right]^I p(1) p(2) n(3) n(4)
\right\rangle ~.
\end{eqnarray}
Note that
\begin{eqnarray}
t(1)\cdot t(2) & = & 1/4 \hspace{1cm} \text{for $T=1$,} \nonumber \\
 & = & -3/4 \hspace{.7cm} \text{for $T=0$,} \nonumber
\end{eqnarray}
and $\langle T=0|V|T=1\rangle=0$, i.e., $\sum_{i<j} t(i) \cdot t(j)$ conserves
isospin.

Skipping over many details, we obtain the expression
\begin{equation}
\langle [J_P J_N]^I | V | [J_P J_N]^I \rangle = 6a - \frac{b}{2} \left[ 1 + 4
(2J_P+1) (2J_N+1) 
\begin{Bmatrix} j & j & J_P \\ j & j & J_N \\ J_P & J_N & I \end{Bmatrix}
\right] ~,
\end{equation}
where the expression on the right contains a $9j$ symbol.

If we sum over $J_P,J_N$ (both even) with the condition $|J_P-J_N|\le I \le J_P
+J_N$, we can equate this to the sum of eigenvalues obtained from the isospin 
formula: 
\begin{equation} 
\sum_{J_P J_N} 6a + \frac{b}{2} \sum_{J_P J_N} \left[
T(T+1) - 3 \right] =  \sum_{J_P J_N} 6a - b \left[ \sum_{J_P J_N} \frac{1}{2} +
X(I) \right] ~,  \label{iso-am} \end{equation} where  \begin{equation} X(I) = 2
\mathop{\sum_{J_P, J_N}}_\text{\footnotesize even} (2J_P +1) (2J_N +1)
\begin{Bmatrix} j & j & J_P \\ j & j & J_N \\ J_P & J_N & I \end{Bmatrix} ~.
\end{equation} 
It should be noted that Zhao and Arima, in considering the
problem of the number of states of angular momentum $I$ for four identical
nucleons in a single $j$ shell, obtained expressions for the values of
$X(I)$~\cite{za05} which display a modular behaviour.

Note that the $a$ terms in Eq.~\eqref{iso-am} cancel out. Then, since every 
remaining term is linear in $b$, it will factor out
\begin{equation}
\frac{1}{2} \sum_{J_P J_N} T(T+1) - \sum_{J_P J_N} \frac{3}{2} = 
-\sum_{J_P J_N} \frac{1}{2} - X(I).
\end{equation}

Let $N(T)$ be the number of states with isospin $T$ ($T=0,1$, and 2). Then, we
get
\begin{equation}
\frac{1}{2} [2N(1)+6N(2)] - \frac{3}{2} [N(0)+N(1)+N(2)] = 
-\frac{1}{2} [N(0)+N(1)+N(2)] - X(I) ~.
\end{equation}
Hence
\begin{equation}
X(I) = N(0) - 2 N(2) ~,
\end{equation}
or, in words,
\begin{eqnarray}
\text{(\# of $T=0$ states)} & - & 2 \text{(\# of $T=2$ states)} \nonumber \\
 & & = 2 \mathop{\sum_{J_P, J_N}}_{\text{\footnotesize even}} (2J_P+1) (2J_N+1)
\begin{Bmatrix} j & j & J_P \\ j & j & J_N \\ J_P & J_N & I \end{Bmatrix} ~.
\label{t0-t2}
\end{eqnarray}

\section{Closing remarks}

Following the initial work of Ginocchio and Haxton~\cite{gh93}, there has been
much recent activity concerning the problem of the number of states of a given
angular momentum $I$~\cite{ze05,za05,rr03,zagy03,za03,za04,za05prc,t05,ze05gh}
for identical fermions in a single $j$ shell. If one limits oneself to
identical particles, e.g., electrons or neutrons, then isospin would seem a
priori to be a useless concept. Secondly, if one has a Hamiltonian which is
simply a constant times the unit matrix, one should not be able to get any
useful information. However, we have shown that if we enlarge the vector space
to include non-identical particles, e.g., neutrons and protons, we can get
useful results for identical particles.

Since two identical nucleons necessarily have isospin $T=1$, then the
interaction $a+b t(1) \cdot t(2)$ is equal to $a + b/4$, i.e., it is a
constant. However, in the larger space of both neutrons and protons, it is no
longer a constant. We took advantage of this fact to get the useful result of
Eq.~\eqref{tracejp}, which had been previously derived by Rosensteel and
Rowe~\cite{rr03} using quasispin considerations for a system of identical
particles, and by Zhao and Arima~\cite{za03} using a $J$-pairing Hamiltonian
also for particles of one kind.

Stimulated by recent work on a system of four identical particles by Zhao and
Arima~\cite{za05prc}, the main emphasis of this work is on a 4-particle system
of both protons and neutrons. Also we go off in a somewhat different direction.
We ask for the number of states of a given angular momentum and a given
isospin. We get a relation which does not answer all the questions, but which
yields the useful result of Eq.~\eqref{t0-t2} for the number of $T=0$ states
minus twice the number of $T=2$ states. Also it is easy to obtain the total
number of states by simply listing the two-particle configurations $J_P,J_N$
that can couple to $I$. We do not bother to give analytic formulas for these.
There are no problems here because each $(J_P,v_P)$ and $(J_N, v_N)$ state
occurs only once.

\begin{acknowledgments}
This work was supported by the U.S. Dept. of Energy under Grant No.
DE-FG0105ER05-02. A.E. is supported by a grant financed by the Secretar\'{\i}a
de Estado de Educaci\'on y Universidades (Spain) and cofinanced by the European
Social Fund.
\end{acknowledgments}

\end{document}